\documentclass[aps,pra,preprint]{revtex4}

\usepackage{amssymb}
\usepackage{graphicx}

\begin{document}
 
 \title[High Temperature Droplet Epitaxy]{High--temperature droplet epitaxy of symmetric GaAs/AlGaAs quantum dots}

\author{Sergio Bietti}
\email{sergio.bietti@unimib.it}
\affiliation{L--NESS and Dipartimento di Scienza dei Materiali, Universit\`a di Milano-Bicocca, via Cozzi 53, I-20125 Milano, Italy} 
\author{Francesco Basso Basset}
\altaffiliation[Now at ]{Dipartimento di Fisica, Sapienza Universit\`a di Roma, Piazzale A. Moro 5, I-00185 Roma, Italy}
\affiliation{L--NESS and Dipartimento di Scienza dei Materiali, Universit\`a di Milano-Bicocca, via Cozzi 53, I-20125 Milano, Italy}
\author{Artur Tuktamyshev}
\affiliation{L--NESS and Dipartimento di Scienza dei Materiali, Universit\`a di Milano-Bicocca, via Cozzi 53, I-20125 Milano, Italy} 
\author{Emiliano Bonera}
\affiliation{L--NESS and Dipartimento di Scienza dei Materiali, Universit\`a di Milano-Bicocca, via Cozzi 53, I-20125 Milano, Italy} 
\author{Alexey Fedorov}
\affiliation{CNR--IFN and L--NESS, via Anzani 42, I--22100 Como, Italy}
\author{Stefano Sanguinetti}
\affiliation{L--NESS and Dipartimento di Scienza dei Materiali, Universit\`a di Milano-Bicocca, via Cozzi 53, I-20125 Milano, Italy} 

\begin{abstract}
We introduce a high--temperature droplet epitaxy procedure, based on the control of the arsenization dynamics of nanoscale droplets of liquid Ga on GaAs(111)A surfaces. The use of high temperatures for the self-assembly of droplet epitaxy quantum dots solves major issues related to material defects, introduced during the droplet epitaxy fabrication process, which limited its use for single and entangled photon sources for quantum photonics applications. We identify the region in the parameter space which allows quantum dots to self--assemble with the desired emission wavelength and highly symmetric shape while maintaining a high optical quality. The role of the growth parameters during the droplet arsenization is discussed and modelled.
\end{abstract}

\maketitle

\section{Introduction}
\label{sec:Intro}

The fabrication of high purity single and entangled photon sources is crucial for the development of quantum communication protocols \cite{KIMBLE08,ORIDIA16} and quantum computation \cite{LJF10,Castelvecchi17}, and it is a fundamental requirement for the realization of repeaters capable of transferring quantum entanglement over long distances \cite{MAT15,WEH18}.  
Among the different light emitting platforms, semiconductor quantum dots (QDs) are very attractive, as they can be integrated with other photonic and electronic components in miniaturized chips. Single photon and entangled photon emitters have been fabricated by QDs using self-assembly techniques like Stranski-Krastanov and Droplet Epitaxy (DE)\cite{PDM19}. In particular, DE enables a fine tuning of the shape, size, density, and thus, of the emission wavelength of the nanostructures \cite{WKG00,MANKOG05,SBS09,SBK10-APL2} with an emission range extending from 700 nm to 1.5 $\mu$m \cite{HLM14}. 
%In particular, QDs by DE were successfully used to realize high quality QDs with vanishing fine structure splitting on GaAs(001) surface \cite{TMW16,HRH17}. 

The high symmetry (111) surface, due to its C$_{3v}$ symmetry, is optimal to reduce the fine structure splitting (FSS) \cite{SINBES09,SWL09,JMA12}, but problematic for the growth via Stranski-Krastanov growth mode, since on (111) the relaxation of a strained III-V semiconductor epilayer immediately proceeds through the nucleation of misfit dislocation at the interface rather than through the formation of coherent 3D islands \cite{YFJ96}. DE is able to self--assemble highly symmetric QDs on (111)A substrate, capable of polarization-entangled photon emission with very high fidelity \cite{KMH13}. Moreover, the choice of GaAs QDs allows a fast radiative recombination and a weaker impact of spin dephasing mechanisms \cite{SSR05,TSH12,JDM13,YSB13}. 

DE QDs have been demonstrated \cite{BBR18} to improve the yield of entanglement--ready photon sources up to 95\% while matching the emission wavelength with an atomic-based optical slow medium such as Rb atoms (as proposed in \cite{AWR11}) for the fabrication of quantum memories and quantum repeaters. This result was achieved through the combination of low values of the excitonic FSS and radiative lifetime, together with the reduced exciton dephasing allowed by the choice of GaAs/AlGaAs QDs fabricated on (111)A substrates. The major DE drawback, related to the low temperature kept for the nanostructure crystallization and barrier layer deposition necessary for this growth technique, was overcome using an innovative high--temperature DE technique which is allowed by the use of (111) substrates.

Here we present and investigate in details a novel high--temperature DE growth technique, which is based on the control of the growth parameters (substrate temperature and As flux) on the Ga adatom diffusion during the GaAs QDs formation by droplet epitaxy on (111)A. In particular, we address the effect of Ga droplet arsenization for the formation of QDs at substrate temperature increased by about 300$^\circ$C with respect to previous reports \cite{MAK10} while avoiding QD elongation \cite{JMS10b} due to anisotropy in Ga adatom diffusion \cite{SBF10NRL2,ABS13,WHM06}. We also address shape control issues, preserving hexagonal shape even at high temperatures, which has a strong impact on the optical quality and excitonic FSS. The highly symmetric dots obtained with our modified recipe show a mean line width of the neutral exciton of about 15 $\mu$eV and a best value of 9 $\mu$eV, a mean fine structure splitting of 4.5 $\mu$eV, which results in the aforementioned large fraction (more than 95\%) of emitters capable of generating entagled photon, as reported in \cite{BBR18}.

\section{Experimental Details}
\label{sec:ExperimentalDetails}
The growth experiments were performed in a conventional Gen II MBE system, on epiready GaAs (111)A substrates. The optimal control of the As flux during the growth was assured by a valved cell. The cracking zone temperature of the As cell was set in every experiment at 600$^\circ$C in order to provide As$_4$ molecules. %Sample growth experiments were monitored \textit{in--situ} by RHEED. 

After the oxide desorption at 580 $^\circ$C, an atomically smooth surface was prepared by growing a 100 nm thick GaAs buffer layer and a 50 nm Al$_{0.3}$Ga$_{0.7}$As barrier layer after reducing the temperature to 520$^\circ$C. To achieve a smooth surface with minimal surface roughness (RMS below 0.5 nm), growth conditions were kept according to \cite{EBF17}. The RHEED pattern clearly showed a (2$\times$2) surface reconstruction \cite{ONK01}.

The substrate temperature was then decreased to 450$^\circ$C and the As valve closed in order to deplete the growth chamber from the arsenic molecules. When the background pressure reached a value below $1\times 10^{-9}$ Torr, a Ga flux with a rate of 0.01 ML/s was supplied to the substrate surface to form Ga droplets. During the Ga supply the surface reconstruction did not show any change. One sample with Ga droplets, from now on D, was then removed from the growth chamber.

For the other samples, in order to study the influence of substrate temperature and As flux during the crystallization on the QD formation, the substrate temperature was decreased to low  temperature, 200 $^\circ$C (sample L1) or medium temperature, 400 $^\circ$C (sample M1) or increased to high temperature, 500$^\circ$C (all the samples of H series) and then the Ga droplets irradiated with As flux for 5 minutes. The irradiated As beam equivalent pressure (BEP) is reported in table \ref{tab:parameters} with all the parameters used for the QD formation. 

\begin{table}
	\centering
		\begin{tabular}{|c||c|c||c|c|c|}
			\hline Sample & Substrate temp.  & Ga amount & Substrate temp. & As BEP & GaAs volume  \tabularnewline & during Ga deposition &  MLs & during arsenization  & Torr &  ($nm^3$/$\mu m^2$) \tabularnewline & ($^\circ$C) &   &  ($^\circ$C) &  &    \\
	\hline\hline D	& 450 & 0.4 & - & - & - \\
	\hline L1	& 450 & 0.4 & 200	& $2\times 10^{-6}$ & $1.15\times10^5$ \\
	\hline M1	& 450 & 0.4 & 400	& $2\times 10^{-6}$ & $6.2\times10^4$ \\
	\hline H0	& 450 & 0.4 & 500	& $8\times 10^{-7}$ & $8.21\times10^3$ \\
	\hline H1	& 450 & 0.4 & 500	& $2\times 10^{-6}$ & $1.03\times10^4$ \\
	\hline H2	& 450 & 0.4 & 500	& $5\times 10^{-6}$ & $2.07\times10^4$ \\
	\hline H3	& 450 & 0.4 & 500	& $2\times 10^{-5}$ & $3.4\times10^4$ \\
	\hline H4	& 450 & 0.4 & 500	& $5\times 10^{-5}$ & $5.6\times10^4$ \\
	\hline H5	& 450 & 0.4 & 500	& $7\times 10^{-5}$ & $5.72\times10^4$ \\
 \hline
		\end{tabular}
	\caption{Substrate temperature and Ga flux of fabricated samples for the droplet formation, substrate temperature and As flux for droplet crystallization, final amount of GaAs measured per square micrometer.}
	\label{tab:parameters}
\end{table}

A second set of samples was then prepared using the same recipe of the first set, but capping the nanostructures with a 10 nm of AlGaAs barrier layer grown at 500 $^\circ$C, another 40 nm at 520 $^\circ$C and a GaAs capping layer of 5 nm. This second set will be recognized adding a C at the end of the sample name.

The morphological characterization of the uncapped samples was performed \textit{ex--situ} by an Atomic Force Microscope (AFM) in tapping mode, using ultra-sharp tips capable of a resolution of about 2 nm. Ensemble photoluminescence measurements (PL) were carried out cooling the samples at 15 K and using the 532 nm line of a Nd:YAG continuous wave laser. The incident power on the samples was 0.5 mW with a laser spot size of approximately 80 $\mu$m.

\section{Results}
\label{sec:Results}
\begin{figure}[h!]
	\centering
		\includegraphics[width=1\textwidth]{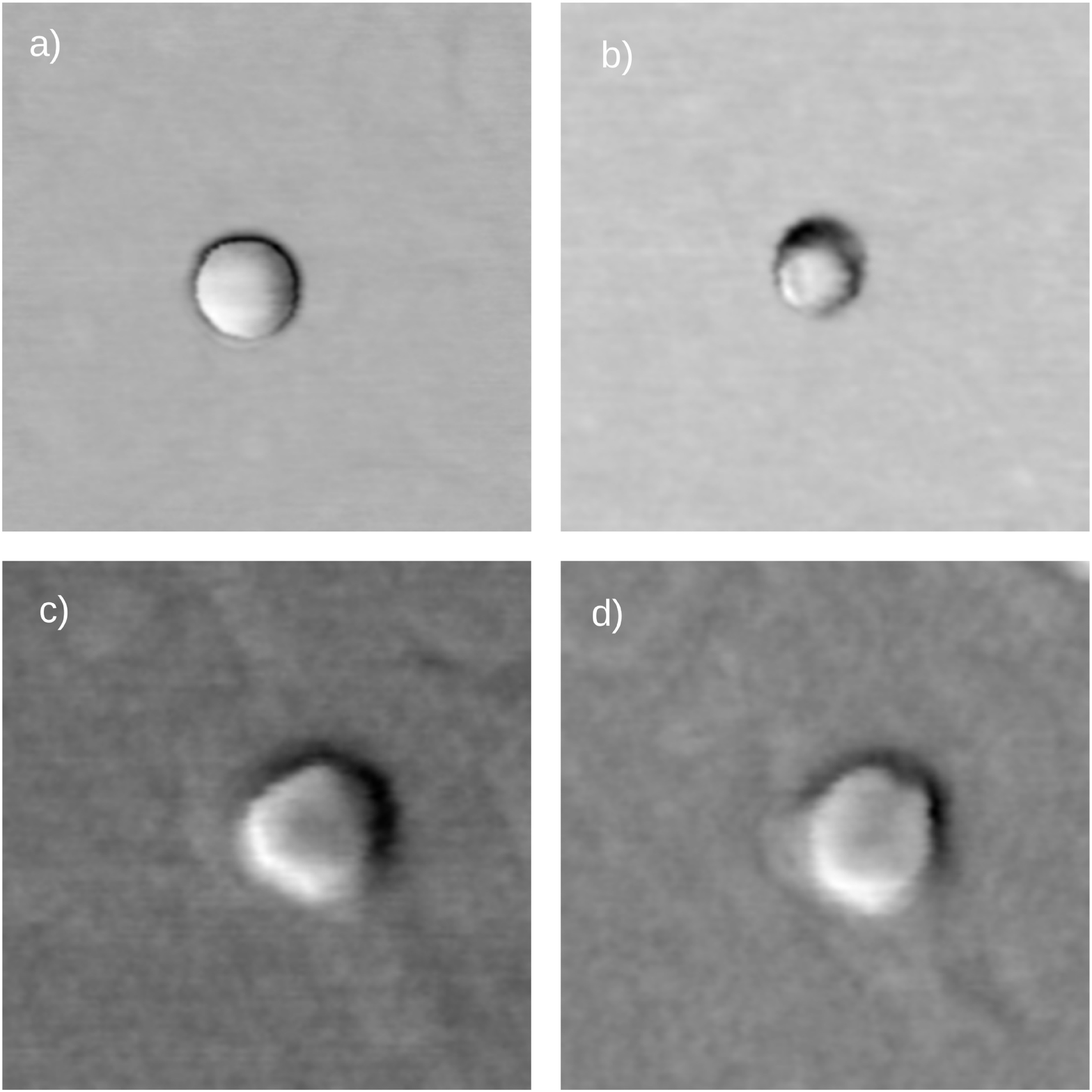}
	\caption{Panel a-d: $250\times250$ nm$^{2}$ AFM scans of sample D (panel a), L1 (panel b), M1 (panel c) and H4 (panel d).}
	\label{fig:sampleBC}
\end{figure}

The AFM characterization of sample D, on which only Ga deposition was performed, shows the formation of Ga droplets with spherical cup shape (see panel a of figure \ref{fig:sampleBC}). The density of the droplets is approximately $7\times 10^{8}$ cm$^{-2}$, the diameter is 50.4 $\pm$ 7.0 nm and the height 7.4 $\pm$ 1.1 nm, the contact angle is approximately 33.7$^\circ$. The shape of the droplet is perfectly symmetric, without elongation in any crystallographic direction. Simple calculations considering the volume of the droplets, the density, and the amount of the deposited Ga, demonstrate with good agreement that all the gallium is collected inside the droplets. This is in agreement with the fact that (111)A surface is Ga terminated and the Ga excess, during gallium deposition, immediately creates droplets on the surface.

The AFM characterization of samples L1, M1 (figure \ref{fig:sampleBC}, panel b and c respectively) and H1 (not shown here), on which droplets were arsenized with a BEP of As of $2\times 10^{-6}$ Torr at a substrate temperature of 200, 400 and 500 $^\circ$C respectively, shows the formation of QDs with different shapes. On sample L1 (see panel b of figure \ref{fig:sampleBC}) truncated pyramids with regular hexagonal base are formed, while truncated pyramids with equilateral triangular base are formed on H1 sample. These shapes are in agreement with the ones reported for larger islands by Jo et al. \cite{JMA12}. On sample H1 the truncated pyramids show a mean height of 2.0 $\pm$ 0.4 nm and the equilateral triangle a mean side of 82.1 $\pm$ 8.6 nm, while on sample L1 the truncated pyramids show a mean height of 10.8 $\pm$ 2.7 nm and the hexagon a mean side of 29.6 $\pm$ 3.3 nm. The angle between the substrate and the sidewalls is approximately 35$^\circ$ for sample L1 and 7$^\circ$ for sample H1. Sample M1 (see panel c of figure \ref{fig:sampleBC}) shows an intermediate behavior. The QDs show a truncated pyramidal shape with an irregular hexagonal base. In this case, three sides are longer than the other three (40.1 $\pm$ 4.2 and 23.8 $\pm$ 3.0 nm, respectively), the mean height 3.0 $\pm$ 0.5 nm  and the angle between the substrate and the sidewalls is approximately 18$^\circ$. 

Sample H4, on which the droplets were arsenized at 500$^\circ$C with a BEP of $5\times 10^{-5}$ torr is shown in figure \ref{fig:sampleBC}d . Here the QDs show a truncated pyramidal shape with regular hexagonal base. The mean height is 3.9 $\pm$ 0.5 nm, the hexagon side of 42.7 $\pm$ 3.2 nm, and the angle between the substrate and the sidewalls is approximately 14$^\circ$. 

In order to study the role of Ga adatom diffusion and incorporation  during droplet arsenization, we analyzed the AFM images and measured the total volume of GaAs crystallized inside the QDs after the arsenization, as reported in table \ref{tab:parameters}. An important parameter of the DE--QDs which helps to elucidate the actual processes during the droplet crystallization is the ratio $\gamma =V_1/V$ between the volume of the final QDs ($V_1$) and the GaAs volume available ($V$) by the complete crystallization of the Ga contained in the droplets. It quantitatively sets the difference between a two--dimensional growth, where the droplet have the role of local group III reservoirs \cite {KOH91}  and $\gamma=0$, and the three--dimensional crystallization of the QD inside the original droplet. The experimental dependence of $\gamma$ on As flux ($J_{\rm As}$) and crystallization temperature $T$ is reported in Figure \ref{fig:gamma}. The measured GaAs crystallized inside the QDs is always lower from the expected volume considering the initial Ga volume stored in the droplets, except for sample L1.  The total volume of GaAs crystallized inside the QDs decreases with increasing arsenization temperature and with decreasing $ J_{\rm As}$. 

Another important fabrication step, which affects the optical properties of the QDs, is the capping procedure. We used ensemble PL to investigate samples L1C, H1C and H4C as shown in figure \ref{fig:PL} in black, red and blue, respectively. The peak around either 640 or 650 nm is related to AlGaAs barrier, whereas the emission bands at 650--690, 660--705 and 700--765 nm, for samples L1C, H1C and H4C respectively, are related to QD emission. The distribution of the emission energy shows sizable modulations which are typical of QDs with low aspect ratio and can be attributed to monolayer fluctuations in height \cite{MAK10}.

%\begin{figure}[h!]
%	\centering
%		\includegraphics[width=1\textwidth]{fig2.eps}
%	\caption{Distribution of QD height and equivalent base radius on sample H4.}
%	\label{fig:distribution}
%\end{figure}

\section{Discussion}
\label{sec:Discussion}
The morphology observed for the QDs grown on the sample on which Ga droplets were arsenized with low As flux, L1, M1 and H1, is in agreement with the model proposed by Jo et al. in \cite{JMA12}. They attributed the morphological evolution to the higher incorporation rate of Ga at A steps (facing the [2$\overline{11}$], [$\overline{1}$2$\overline{1}$] and [$\overline{11}$2] directions), respect to that at the B steps (facing the [$\overline{2}$11], [1$\overline{2}$1] and [11$\overline{2}$]). The triangular shape observed for the QDs crystallized at low As flux and high substrate temperature can be consequently explained as an effect of the different incorporation rates of Ga adatoms on the two steps. 

\begin{figure}[h!]
	\centering
		\includegraphics[width=1\textwidth]{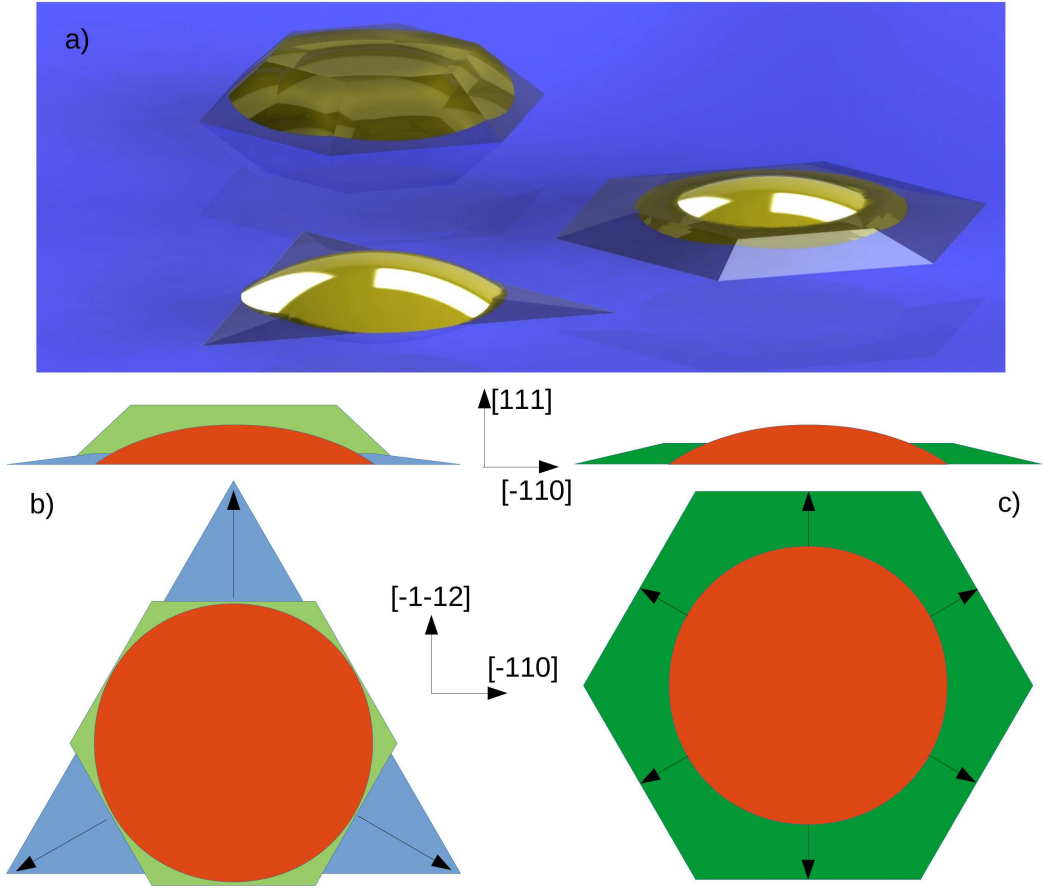}
	\caption{Panel a: sketch of mean sized quantum dot on sample L1, H1 and H4 (in brown, respectively upper, lower left and lower right) compared with the original gallium droplet (yellow). Panel b: graphical representation of mean sized droplet on sample D (orange) and of quantum dot on samples H1 (light blue) and L1 (light green). Panel c: graphical representation of mean sized droplet on sample D (orange) and of quantum dot on samples H4 (dark green).}
	\label{fig:graphcomp}
\end{figure}

Let us analyze more in the details the change of the QD morphology as a function of the growth parameters. The shape of the QDs on samples L1 and H1 are graphically summarized in figure \ref{fig:graphcomp} and compared with the size of the original droplet. Comparing in panel b) the mean dimensions of Ga droplets on sample D (orange), and of QDs on samples L1 (light green) and H1 (light blue), it is possible to see that the formation of QD crystallized with $2\times 10^{-6}$ torr As BEP, while increasing the substrate temperature from 200 to 500 $^\circ$C, is dominated by incorporation exclusively along A steps, while the incorporation along B steps is suppressed. It can be observed clearly from panel b of fig. \ref{fig:graphcomp} that the sides of the hexagonal dot on sample L1 (in light green) is mostly tangent to the base circle of the original droplet (orange), and that the sides of triangular dots on sample H1 (light blue) are mostly tangent to the base circle of the original droplet only along B steps, thus confirming the absence of incorporation in those directions. Also on sample M1 (not shown in the picture) the longer sides are mostly tangent along B steps to the base circle of the original droplet. 

These observations confirm that the shape of the QDs is determined by kinetically controlled diffusion and incorporation processes in which the different growth velocity between A and B steps determines anisotropy at high temperature. 

By increasing the As BEP up to $5\times 10^{-5}$ torr at high substrate temperature, it is possible to obtain again QDs with hexagonal symmetrical shape. A graphical representation is reported in panel c of figure \ref{fig:graphcomp}. Here we consider the shape and the mean size measured for the original gallium droplet (orange) and we compared it with the shape and the mean size measured for the QDs on sample H4 (dark green). From panel c can also be observed that on sample H4 all the sides of the hexagonal dots are away but at the same distance from the base circle of the original droplet. This means that in these arsenization conditions the incorporation of Ga adatoms along A and  B steps, has comparable speeds. 

The described behaviour allows the growth of GaAs QDs by DE at substrate temperature much higher than the one typically used on (001) substrates and, compared to the data previously reported on (111)A substrates, to preserve the hexagonal shape also for arsenization performed up to 500$^\circ$C. This is expected to allow for the growth of materials with improved crystalline quality respect to the usual DE QDs crystallized at 200$^\circ$C. In fact, a low temperature of crystallization for the Ga droplets \cite{SWK02,SMG08} and subsequent AlGaAs barrier deposition \cite{KKW10} is detrimental for the crystalline and the optical quality of the QDs, mainly due to the formation to a high density of vacancies and the incorporation of defects. 

\begin{figure}[h!]
	\centering
		\includegraphics[width=1\textwidth]{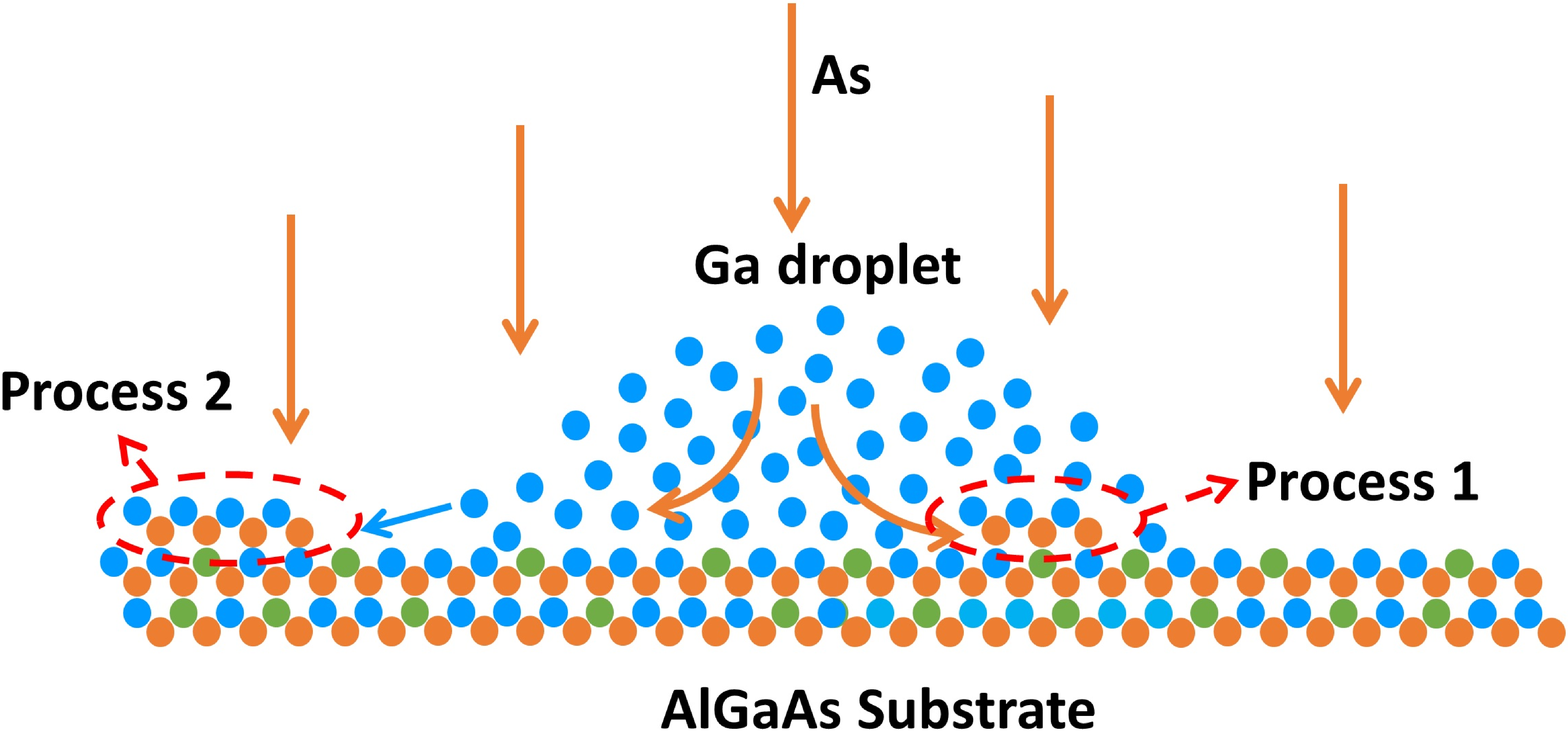}
	\caption{Schematics of the growth processes active during DE. Process 1 refers to As incorporation at the bottom of the droplet. Process 2 refers to Ga atom detachment from the droplet and subsequent incorporation into the crystal via As reaction}
	\label{fig:Model}
\end{figure}

To understand the reasons beyond the observed behavior, we have to consider the control of the growth kinetic, which allows to tune the fabricated nanostructures from three to two--dimensional. 
On GaAs (001) substrates, nanostructure shape can vary from compact islands to rings and eventually to flat disks extending to the droplet surroundings \cite{SBS12}. These different shapes can be achieved by tuning the speed of the arsenization processes that takes place within the metallic droplet (process 1 in Figure \ref{fig:Model}), and the Ga atoms diffusion and incorporation outside the droplet (process 2 in Figure \ref{fig:Model}). 

% On GaAs (001) surface, a growth dominated by the first mechanism leads to the formation of QDs \cite{BSS13} while favoring the diffusion mechanism leads to composite nanostructures extended outside the rim of the original droplet. On (001) surface, the balance between the two mechanisms is determined by $T$ and  $ J_{\rm As}$ \cite{BSA09}. In both cases, all the initial amount of liquid Ga is crystallized within the final GaAs nanostructure \cite{ABS13}.

On GaAs (111)A substrates, the geometry of the fabricated nanostructure is different due to different symmetry, and we observe compact islands for all the measured parameter range. %but, also on this surface, we observe that the ratio of Ga crystallized inside the rim of the original droplet and the one outside, is changing with the growth parameters $T$ and  $ J_{\rm As}$.

The parameter which quantifies the balance between process 1 (the crystallization inside the original rim of the droplet) and process 2 (diffusion/incorporation) is the ratio $\gamma$, as it is expected to range from one in the pure three--dimensional growth to zero in the diffusion/incorporation two--dimensional growth. As previously reported for (001) surface, $\gamma$ is a function of $T$ and  $ J_{\rm As}$ \cite{BSE14}.

%A major difference between the (001) and the (111)A surface can then be shown by considering the ratio $\gamma =V_1/V$ between the actual volume of the QDs after the arsenization ($V_1$) and the volume of GaAs achievable by the complete crystallization of the Ga contained in the initial droplets ($V$).

%On the (001) surface the final volume of the nanostructures $V_1$ and the volume of the total Ga available after arsenization coincide also in the case of incorporation outside the rim of the original droplets \cite{ABS13}, while on (111)A surface the possibility of a loss of material outside the final nanostructure was already observed for InAs QDs grown by DE \cite{BEF15}.

\begin{figure}[h!]
	\centering
		\includegraphics[width=1\textwidth]{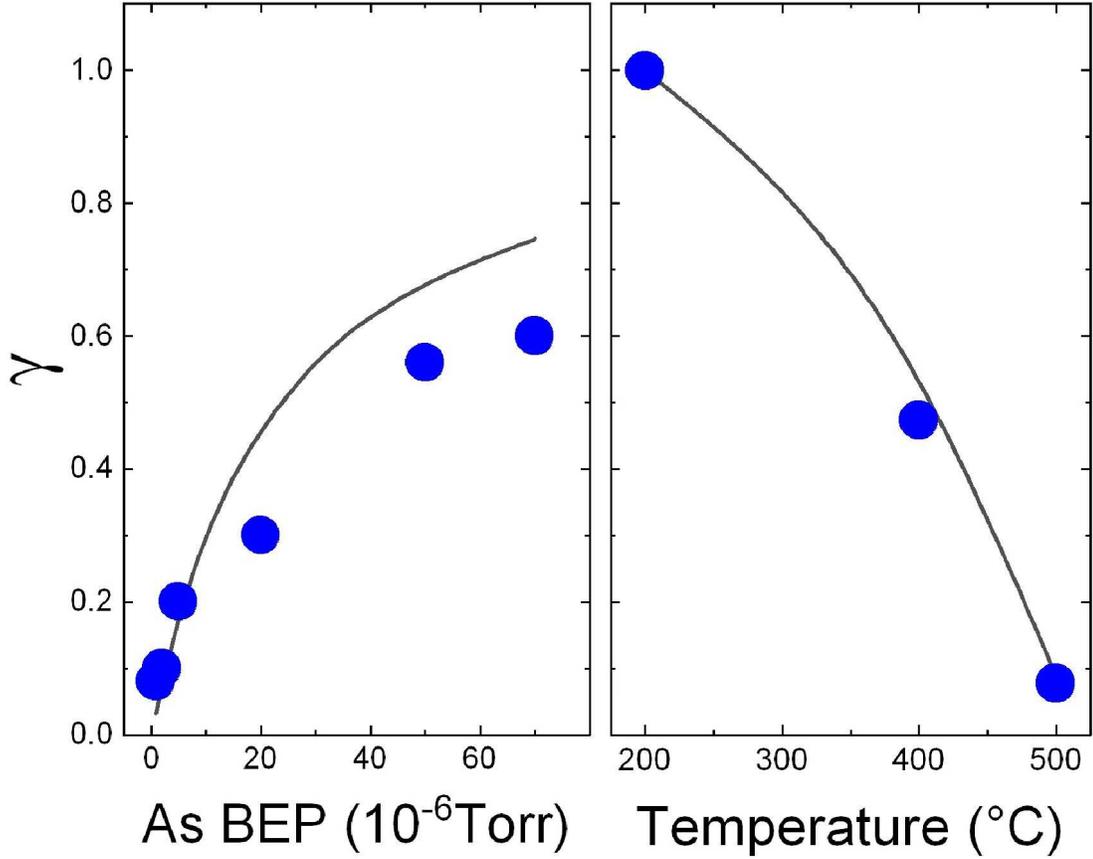}
	\caption{Dependence of the ratio $\gamma = V_1/V$ on the As BEP $J_{\rm As}$ (left panel) and substrate temperature $T$ (right panel). The experimental data are indicated by the red circles. The continous line reports the fit of the data using Eq. (\ref{eq:gamma}). }
	\label{fig:gamma}
\end{figure}

In figure \ref{fig:gamma} we reported the data related to parameter $\gamma$ for the temperature series (L1, M1 and H1, from the top down in right panel) and for the As pressure series (from H0 to H5, from left to right, left panel). Considering the temperature series, it is possible to see that on sample L1 almost all the gallium is crystallized inside the hexagonal QDs, while on samples M1 and H1 a loss of about 46\% and 91\% of the original gallium deposited is measured. 
Considering the samples of As pressure series, arsenized at 500$^\circ$ C with different As fluxes from $8\times10^{-7}$ to $7\times10^{-5}$ Torr, it is also consistently observed a loss in volume. The value of $\gamma$ is increasing with the equivalent pressure of As irradiated during the droplet crystallization. For this series the loss in volume of GaAs crystallized inside the QDs is between 50\% and 93\%.
It is interesting to notice that, on (111)A surface, the possibility of a loss of material outside the final nanostructure was already observed for InAs QDs grown by DE in \cite{BEF15}. 

To understand the observed behavior we have to consider how the droplet arsenization and the diffusion processes depend on the growth parameters.
Let us first assume the droplet crystallization mechanism depends on the liquid--solid interface area, the arsenic solubility and diffusivity into the droplet and $J_{\rm As}$. Considering, as first order approximation, a slow dependence of the interface area on the growth time (process 1 in figure \ref{fig:Model}, the volume crystallized inside the droplet depends linearly on growth time $t$ via the equation 
\begin{equation}V_1(t) = \rho_D J_{\rm As} t
\label{eq:V1}
\end{equation} 
where $\rho_D$ is the constant that takes into account all the other factors (the arsenic solubility and diffusivity into the droplet) with the exception of the As flux $J_{\rm As}$. %There is a limit in the As incorporation probability at high flux, due to droplet saturation, which makes $V_1$ to deviate from the  $ J_{\rm As}$ proportionality. However, actual experimental conditions have been demonstrated not to limit growth rate provided that the liquid/solid interface remains constant \cite{BSS13,Reyes2013}. 

The growth rate of Ga adatoms incorporated outside the nanostructure depends in a more complex way on $J_{\rm As}$ and $T$. The easiest resulting geometry observed on GaAs(001) substrates is a disk, with a radius given by the sum of the diffusion length of Ga atoms ($\ell$) and of the radius of the droplet: $R = \ell(J_{\rm As},T) + r_0$ \cite{BSE14}. Here $\ell^2=D_0 \exp (-E_D/kT) (N_d/J_{\rm As})$, where $D_0$ is the diffusivity prefactor, $E_D$ the diffusion activation energy and $N_d$ the surface density of As sites. The disk is therefore increasing its radius by increasing the substrate temperature and decreasing the As flux. The disk increases its thickness with a rate which is proportional to $J_{\rm As}$ and to the product between As/Ga reaction probability and  As residence time  $\zeta_R$ \cite{ZZT13}. Considering that in most of the cases $\ell \gg r_0$ (with the exception of the low T range) and that the diffusion/incorporation process follows the same physics on (001) an (111)A, the growth rate of process 2 can be expressed by 
\begin{eqnarray}
V_2\left(t\right)  &=&  \mu' \zeta_R \ell^2 J_{As}t=\mu \zeta_R \left[D_0 \exp (-E_D/kT)  {J_{As}^{-1}}\right]J_{As}t \nonumber\\
 &=&\mu \zeta_R D_0 \exp (-E_D/kT) t
\label{eq:V2}
\end{eqnarray}
Where $\mu'$ and $\mu$ are constant collecting geometrical and constant factors.
The growth will proceed for a time $\tau$ until the Ga in the droplet is fully consumed. This condition is reached when $V = V_1(\tau)+V_2(\tau)$. Combining Eq. (\ref{eq:V2})  equation with Eq (\ref{eq:V1}) 
\begin{equation}
\gamma(J_{\rm As},T) = \left[ {1 + \frac{\mu \zeta_R D_0 \exp(-E_D/kT)}{\rho_D J_{\rm As}}} \right]^{-1}
\label{eq:gamma}
\end{equation}
The dependence of $\gamma$ on $J_{\rm As}$ and $T$ is reported in Figure \ref{fig:gamma} and compared with experimental values of the ratio. For the diffusion activation energy on GaAs (111)A surface we used the value as calculated by Ref. \cite{SLH11}, $E_D$=1.06 eV. The value of the ratio $\mu \zeta_R D_0/\rho_D = 2\times 10^2$ Torr has been fitted to the temperature series L1, M1, and H1. The data are nicely reproduced by our model which depends on a single fit parameter. It is interesting to notice that increasing the term $\mu \zeta_R D_0/\rho_D$, we decrease the value of $\gamma$. For this reason, a short As residence time $\zeta_R$ on (111)A respect to (001) substrates, as reported in \cite{SFJ94}, is expected to make the contribution of the crystallization process inside the droplet to be preminent even at high T, provided a sufficient increase of the As BEP. The upper limit for this effect is marked by the limited As solubility and diffusivity in the droplet for extremely high As flux.
% and by the possibility to start of a Mullins-Sekerka instability, which produces GaAs islands with cores of metallic unreacted Ga at high As flux. 

\begin{figure}[h!]
	\centering
		\includegraphics[width=1\textwidth]{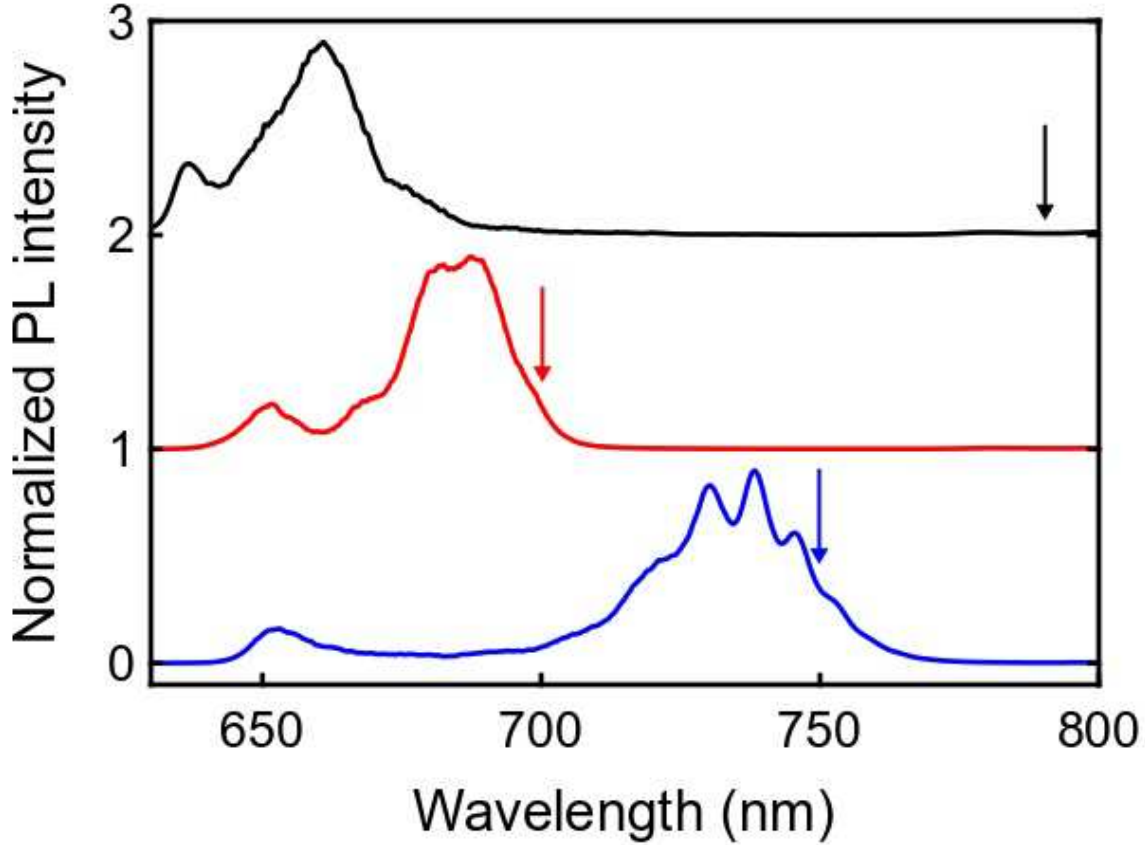}
	\caption{Normalized low-temperature ensemble PL spectra of the samples L1C, H1C and H4C (from top down) capped with a barrier layer. The arrows mark the calculated emission wavelength.}
	\label{fig:PL}
\end{figure}

Understanding the relationship between the growth parameters and the shape of the nanostructures is fundamental for the fabrication of emitters with specific electronic and optical properties but it is then necessary to evaluate the effect of the deposition of the AlGaAs capping layer in terms of shape change and interdiffusion. The optical properties of the QDs capped with 50 nm Al$_{0.3}$Ga$_{0.7}$As and 5 nm of GaAs were studied by means of ensemble photoluminescence (PL) and single-band constant-potential model simulations on the second set of sample. As expected from simple considerations of quantum confinement energy, the size of the QDs, and in particular their height, is affecting the emission wavelength. The ensemble PL spectra of samples L1C, H1C and H4C are displayed in black, red and blue, respectively, in Fig. \ref{fig:PL}. %The peak around either 640 or 650 nm is related to AlGaAs barrier, whereas the emission bands at 650--690, 660--705 and 700--765 nm, for samples L1C, H1C and H4C respectively, are related to QD emission. The distribution of the emission energy shows sizable modulations which are typical of QDs with low aspect ratio and can be attributed to monolayer fluctuations in height \cite{MAK10}.
We simulated the expected radiative recombination energies with the single-band constant-potential model \cite{MARBAS94}, using realistic dot shapes from the AFM images taken from the corresponding uncapped samples and linear dimensions given by average values from the experimental size distribution. The band parameters used in the calculation \cite{KMO05, PAVGUZ94} are chosen consistently with previous studies on droplet epitaxy GaAs/AlGaAs QDs. The results for the ground state transition are shown by arrows in Fig. \ref{fig:PL} alongside the ensemble PL spectra. It is worth notice that for the samples in which the substrate temperature during the QDs crystallization was set equal or higher than 400$^\circ$C, a small blueshift around 30 meV was found between the theoretical estimation and the centroid of the energy distribution. For a substrate temperature of 200$^\circ$C during the As crystallization (sample L1C), a blueshift larger than 200 meV was measured. This large discrepancy is attributed to sizeable interdiffusion at the AlGaAs/GaAs interface during the capping step, when the temperature is substantially increased up to 500$^\circ$C and the nanostructure is slowly covered with an AlGaAs layer. 
%While the interdiffusion processes during the deposition of the capping layer are always present, they have much more limited impact for QDs crystallized at elevated temperature, and can be estimated in an interdiffusion of few monolayers. Also in this case, the effect demonstrates the high crystalline quality of the QDs grown at elevated temperature, as in the growth of GaAs DE--QDs, the enhancement of interdiffusion process can be attributed to the presence of vacancies and crystal defect, as described in \cite{SWK02,SMG08}. 
We observe the presence of interdiffusion during the deposition of the capping layer in all our samples. However only in sample L1, where droplet are crystallized at 200$^\circ$C, the interdiffusion has a strong impact. On the contrary, the QDs crystallized at $T>400\ ^\circ$C show a limited blue--shift. Theoretical predictions simulating QDs with and without interdiffusion at the boundaries, show that this process involves only a few monolayers.  The limited impact of capping demonstrates, again, the high crystalline quality of the QDs grown at elevated temperature. 

%Finally, we used above-barrier excitation micro-PL measurement to study the effect of the thickness of the AlGaAs barrier on the linewidth of the exciton transition in the single QD. We fabricated three different samples embedding hexagonal GaAs QDs in the middle of AlGaAs barrier layer of 100, 200 and 300 nm. For the QD formation on these three samples we used the same parameters reported for sample H4 (i.e. 0.4 MLs of Ga deposited at 450$^\circ$C, arsenization with $3\times10^(-5)$ torr BEP at 500$^\circ$C). The linewidth of the excitonic transition was calculated by extracting the full width at half maximum from a Gaussian fit performed on spectral peaks identified as neutral excitonic transitions. The mean values of linewidth were 170, 60 and 15 $\mu$eV, respectively. This improvement related to the barrier thickness can be explained considering that the transient electrostatic field from surface charges at the vacuum--semiconductor interface is reported to be  a major cause for spectral wandering in GaAs/AlGaAs droplet epitaxy quantum dots grown on (100) substrates  \cite{HMC15}. Only the deposition of a thick enough AlGaAs barrier increases the distance between the QDs and the free surface charges, reducing the spectral wandering and allowing to obtain narrower lines.

\section{Conclusions}
\label{sec:Conclusions}
The presented high temperature droplet epitaxy procedure allows for the self--assembly of QDs with high optical quality and symmetric shape, as we demonstrated in \cite{BBR18} in terms of exciton linewidth (mean value 15 $\mu$eV, best 9 $\mu$eV) and FSS (mean value below 5 $\mu$eV). These properties are fundamental for the fabrication of entanglement--ready photon sources and we demonstrated \cite{BBR18} that 95\% of the emitters can deliver the photon pairs deterministically and with the fidelity to the expected Bell state above the classical limit. The improved optical quality is related to the high crystalline quality of the GaAs QDs and the surrounding AlGaAs barrier, as both are crystallized and deposited at a temperature close to the optimal one for the GaAs crystal growth on (111)A surface. 
We investigated and modeled the dependence of the QD shape and size on the parameters used during the crystallization process. We found that high temperature droplet epitaxy on (111)A substrates is governed, as the standard droplet epitaxy on GaAs(001), by the balance  between crystallization within the droplet and the process of Ga adatom detachment from the droplet, diffusion and incorporation into the crystal surrounding the droplet. The predominance of the former over the latter allows for the self--assembly of 3D islands. This is realized on GaAs (111)A substrates at high T owing to the low residence time of the As on the (111)A surface which hinders the diffusion/crystallization processes on the crystal surface around the droplet. The high As pressure required for the crystallization also permits the equalization of the growth velocities along the A and B steps resulting in a symmetric hexagonal shape is obtained. The temperature of crystallization also permits to preserve the shape of the QDs when capped, thus allowing for the reproducibility of the fabrication procedures which is a fundamental asset for the deterministic design of the emitters for wavelength--specific applications.

The authors acknowledge financial support through ITN 4-Photon Marie Sk\l{}odowska-Curie Grant Agreement No 721394.
\newpage

\bibliographystyle{unsrt}
\bibliography{additional,library}

\begin{thebibliography}{10}

\bibitem{KIMBLE08}
H.J. Kimble.
\newblock The quantum internet.
\newblock {\em Nature}, 453(7198):1023--1030, 2008.
\newblock cited By 1631.

\bibitem{ORIDIA16}
Adeline Orieux and Eleni Diamanti.
\newblock Recent advances on integrated quantum communications.
\newblock {\em Journal of Optics}, 18(8):083002, 2016.

\bibitem{LJF10}
Thaddeus~D. Ladd, Fedor Jelezko, Raymond Laflamme, Yasunobu Nakamura,
  Christopher~R. Monroe, and Jeremy~Lloyd O'Brien.
\newblock Quantum computers.
\newblock {\em Nature}, 464:45--53, 2010.

\bibitem{Castelvecchi17}
Davide Castelvecchi.
\newblock Quantum computers ready to leap out of the lab in 2017.
\newblock {\em Nature}, 541:9--10, 2017.

\bibitem{MAT15}
William~J. Munro, Koji Azuma, Kiyoshi Tamaki, and Kae Nemoto.
\newblock Inside quantum repeaters.
\newblock {\em IEEE Journal of Selected Topics in Quantum Electronics},
  21:78--90, 2015.

\bibitem{WEH18}
Stephanie Wehner, David Elkouss, and Ronald Hanson.
\newblock Quantum internet: A vision for the road ahead.
\newblock {\em Science}, 362(6412), 2018.

\bibitem{PDM19}
Massimo Gurioli, Zhiming Wang, Armando Rastelli, Takashi Kuroda, and Stefano
  Sanguinetti.
\newblock Droplet epitaxy of semiconductor nanostructures for quantum photonic
  devices.
\newblock {\em Nature Materials}, 2019.

\bibitem{WKG00}
Katsuyuki Watanabe, Nobuyuki Koguchi, and Yoshihiko Gotoh.
\newblock {Fabrication of GaAs Quantum Dots by Modified Droplet Epitaxy}.
\newblock {\em Japanese Journal of Applied Physics}, 39(2):L79--L81, 2000.

\bibitem{MANKOG05}
Takaaki Mano and Nobuyuki Koguchi.
\newblock {Nanometer-scale GaAs ring structure grown by droplet epitaxy}.
\newblock {\em Journal of Crystal Growth}, 278(1-4):108--112, may 2005.

\bibitem{SBS09}
Claudio Somaschini, Sergio Bietti, Nobuyuki Koguchi, and Stefano Sanguinetti.
\newblock {Fabrication of multiple concentric nanoring structures}.
\newblock {\em Nano Letters}, 9(10):3419--3424, oct 2009.

\bibitem{SBK10-APL2}
Claudio Somaschini, Sergio Bietti, Nobuyuki Koguchi, and Stefano Sanguinetti.
\newblock {Shape control via surface reconstruction kinetics of droplet epitaxy
  nanostructures}.
\newblock {\em Applied Physics Letters}, 97(20):203109, 2010.

\bibitem{HLM14}
Neul Ha, Xiangming Liu, Takaaki Mano, Takashi Kuroda, Kazutaka Mitsuishi,
  Andrea Castellano, Stefano Sanguinetti, Takeshi Noda, Yoshiki Sakuma, and
  Kazuaki Sakoda.
\newblock {Droplet epitaxial growth of highly symmetric quantum dots emitting
  at telecommunication wavelengths on InP(111)A}.
\newblock {\em Applied Physics Letters}, 104(14):143106, apr 2014.

\bibitem{SINBES09}
Ranber Singh and Gabriel Bester.
\newblock Nanowire quantum dots as an ideal source of entangled photon pairs.
\newblock {\em Phys. Rev. Lett.}, 103:063601, Aug 2009.

\bibitem{SWL09}
Andrei Schliwa, Momme Winkelnkemper, Anatol Lochmann, Erik Stock, and Dieter
  Bimberg.
\newblock In(ga)as/gaas quantum dots grown on a (111) surface as ideal sources
  of entangled photon pairs.
\newblock {\em Phys. Rev. B}, 80:161307, Oct 2009.

\bibitem{JMA12}
Masafumi Jo, Takaaki Mano, Marco Abbarchi, Takashi Kuroda, Yoshiki Sakuma, and
  Kazuaki Sakoda.
\newblock {Self-limiting growth of hexagonal and triangular quantum dots on
  (111)A}.
\newblock {\em Crystal Growth and Design}, 12(111):1411--1415, 2012.

\bibitem{YFJ96}
H.~Yamaguchi, M.~R. Fahy, and B.~a. Joyce.
\newblock {Inhibitions of three dimensional island formation in InAs films
  grown on GaAs (111)A surface by molecular beam epitaxy}.
\newblock {\em Applied Physics Letters}, 69(6):776, 1996.

\bibitem{KMH13}
Takashi Kuroda, Takaaki Mano, Neul Ha, Hideaki Nakajima, Hidekazu Kumano,
  Bernhard Urbaszek, Masafumi Jo, Marco Abbarchi, Yoshiki Sakuma, Kazuaki
  Sakoda, Ikuo Suemune, Xavier Marie, and Thierry Amand.
\newblock Symmetric quantum dots as efficient sources of highly entangled
  photons: Violation of bell's inequality without spectral and temporal
  filtering.
\newblock {\em Phys. Rev. B}, 88:041306, Jul 2013.

\bibitem{SSR05}
R.~Seguin, A.~Schliwa, S.~Rodt, K.~P\"otschke, U.~W. Pohl, and D.~Bimberg.
\newblock Size-dependent fine-structure splitting in self-organized
  $\mathrm{InAs}/\mathrm{GaAs}$ quantum dots.
\newblock {\em Phys. Rev. Lett.}, 95:257402, Dec 2005.

\bibitem{TSH12}
J.~Treu, C.~Schneider, A.~Huggenberger, T.~Braun, S.~Reitzenstein, S.~Höfling,
  and M.~Kamp.
\newblock Substrate orientation dependent fine structure splitting of symmetric
  in(ga)as/gaas quantum dots.
\newblock {\em Applied Physics Letters}, 101(2):022102, 2012.

\bibitem{JDM13}
Gediminas Juska, Valeria Dimastrodonato, Lorenzo~O. Mereni, Agnieszka
  Gocalinska, and Emanuele Pelucchi.
\newblock Towards quantum-dot arrays of entangled photon emitters.
\newblock {\em Nature Photonics}, 7:527 EP --, May 2013.

\bibitem{YSB13}
{Christopher D.} Yerino, {Paul J.} Simmonds, Baolai Liang, Daehwan Jung,
  Christian Schneider, Sebastian Unsleber, Minh Vo, {Diana L.} Huffaker, Sven
  H{\"o}fling, Martin Kamp, and {Minjoo Lawrence} Lee.
\newblock Strain-driven growth of gaas(111) quantum dots with low fine
  structure splitting.
\newblock {\em Applied Physics Letters}, 105(25), 12 2014.

\bibitem{BBR18}
Francesco~Basso Basset, Sergio Bietti, Marcus Reindl, Luca Esposito, Alexey
  Fedorov, Daniel Huber, Armando Rastelli, Emiliano Bonera, Rinaldo Trotta, and
  Stefano Sanguinetti.
\newblock {High-Yield Fabrication of Entangled Photon Emitters for Hybrid
  Quantum Networking Using High-Temperature Droplet Epitaxy}.
\newblock {\em Nano Letters}, 18:505, 2018.

\bibitem{AWR11}
N.~Akopian, L.~Wang, A.~Rastelli, O.~G. Schmidt, and V.~Zwiller.
\newblock Hybrid semiconductor-atomic interface: slowing down single photons
  from a quantum dot.
\newblock {\em Nature Photonics}, 5:230 EP --, Feb 2011.

\bibitem{MAK10}
Takaaki Mano, Marco Abbarchi, Takashi Kuroda, Brian Mcskimming, Akihiro Ohtake,
  Kazutaka Mitsuishi, and Kazuaki Sakoda.
\newblock {Self-assembly of symmetric GaAs quantum dots on (111)A substrates:
  Suppression of fine-structure splitting}.
\newblock {\em Applied Physics Express}, 3(6):065203, may 2010.

\bibitem{JMS10b}
Masafumi Jo, Takaaki Mano, and Kazuaki Sakoda.
\newblock {Unstrained GaAs quantum dashes grown on GaAs(001) substrates by
  droplet epitaxy}.
\newblock {\em Applied Physics Express}, 3(4):045502, apr 2010.

\bibitem{SBF10NRL2}
Claudio Somaschini, Sergio Bietti, a.~Fedorov, Nobuyuki Koguchi, and Stefano
  Sanguinetti.
\newblock {Concentric Multiple Rings by Droplet Epitaxy: Fabrication and Study
  of the Morphological Anisotropy}.
\newblock {\em Nanoscale Research Letters}, 5:1865--1867, aug 2010.

\bibitem{ABS13}
Silvia Adorno, Sergio Bietti, and Stefano Sanguinetti.
\newblock {Annealing induced anisotropy in GaAs/AlGaAs quantum dots grown by
  droplet epitaxy}.
\newblock {\em Journal of Crystal Growth}, 378:515--518, sep 2013.

\bibitem{WHM06}
Zhiming~M. Wang, Kyland Holmes, Yuriy~I. Mazur, Kimberly~A. Ramsey, and
  Gregory~J. Salamo.
\newblock Self-organization of quantum-dot pairs by high-temperature droplet
  epitaxy.
\newblock {\em Nanoscale Research Letters}, 1(1):57, Jul 2006.

\bibitem{EBF17}
Luca Esposito, Sergio Bietti, Alexey Fedorov, Richard Noetzel, and Stefano
  Sanguinetti.
\newblock {Ehrlich-Schwoebel Effect on the Growth Dynamics of GaAs(111)A
  surfaces}.
\newblock {\em PHYSICAL REVIEW MATERIALS}, 1:024602, 2017.

\bibitem{ONK01}
Akihiro Ohtake, Jun Nakamura, Takuji Komura, Takashi Hanada, Takafumi Yao,
  Hiromi Kuramochi, and Masashi Ozeki.
\newblock {Surface structures of GaAs{\{}111{\}}A,B-(2×2)}.
\newblock {\em Physical Review B}, 64(4):045318, jun 2001.

\bibitem{KOH91}
Kiyoshi Kanisawa, Jiro Osaka, Shigeru Hirono, and Naohisa Inoue.
\newblock {Al-Ga monolayer lateral growth observed in situ by scanning electron
  microscopy}.
\newblock {\em Applied Physics Letters}, 58(21):2363--2365, 1991.

\bibitem{SWK02}
Stefano Sanguinetti, Katsuyuki Watanabe, Takashi Kuroda, F.~Minami, Y.~Gotoh,
  and Nobuyuki Koguchi.
\newblock {Effects of post-growth annealing on the optical properties of
  self-assembled GaAs/AlGaAs quantum dots}.
\newblock {\em Journal of Crystal Growth}, 242:321--331, 2002.

\bibitem{SMG08}
Stefano Sanguinetti, Takaaki Mano, a.~Gerosa, Claudio Somaschini, Sergio
  Bietti, Nobuyuki Koguchi, E.~Grilli, M.~Guzzi, Massimo Gurioli, and Marco
  Abbarchi.
\newblock {Rapid thermal annealing effects on self-assembled quantum dot and
  quantum ring structures}.
\newblock {\em Journal of Applied Physics}, 104(11):113519, 2008.

\bibitem{KKW10}
Keiji Kuroda, Takashi Kuroda, Katsuyuki Watanabe, Takaaki Mano, Giyuu Kido,
  Nobuyuki Koguchi, and Kazuaki Sakoda.
\newblock {Distribution of exciton emission linewidth observed for GaAs quantum
  dots grown by droplet epitaxy}.
\newblock {\em Journal of Luminescence}, 130(12):2390--2393, 2010.

\bibitem{SBS12}
Claudio Somaschini, Sergio Bietti, a.~Scaccabarozzi, E.~Grilli, and Stefano
  Sanguinetti.
\newblock {Self-assembly of quantum dot-disk nanostructures via growth kinetics
  control}.
\newblock {\em Crystal Growth and Design}, 12(3):1180--1184, mar 2012.

\bibitem{BSE14}
Sergio Bietti, Claudio Somaschini, Luca Esposito, Alexey Fedorov, and Stefano
  Sanguinetti.
\newblock {Gallium surface diffusion on GaAs (001) surfaces measured by
  crystallization dynamics of Ga droplets}.
\newblock {\em Journal of Applied Physics}, 116(11):114311, sep 2014.

\bibitem{BEF15}
Sergio Bietti, Luca Esposito, Alexey Fedorov, Andrea Ballabio, Andrea
  Martinelli, and Stefano Sanguinetti.
\newblock {Characterization and Effect of Thermal Annealing on InAs Quantum
  Dots Grown by Droplet Epitaxy on GaAs(111)A Substrates}.
\newblock {\em Nanoscale Research Letters}, 10(1):247, 2015.

\bibitem{ZZT13}
Z~Y Zhou, C~X Zheng, W~X Tang, J.~Tersoff, and David~E Jesson.
\newblock {Origin of Quantum Ring Formation During Droplet Epitaxy}.
\newblock {\em Physical Review Letters}, 111(3):036102, jul 2013.

\bibitem{SLH11}
J.~N. Shapiro, A.~Lin, D.~L. Huffaker, and C.~Ratsch.
\newblock {Potential energy surface of in and Ga adatoms above the (111)A and
  (110) surfaces of a GaAs nanopillar}.
\newblock {\em Physical Review B - Condensed Matter and Materials Physics},
  84(8):1--5, 2011.

\bibitem{SFJ94}
M.; Joyce~B. Sato, K.;~Fahy.
\newblock Reflection high energy electron diffraction intensity oscillation
  study of the growth of gaas on gaas(111)a.
\newblock {\em Surface Science}, 315:105--111, 1994.

\bibitem{MARBAS94}
G.~Bastard {J. Y. Marzin}.
\newblock {CALCULATION OF THE ENERGY LEVELS IN InAs/GaAs QUANTUM DOTS}.
\newblock {\em Solid state communications}, 92(5):437--442, 1994.

\bibitem{KMO05}
Takashi Kuroda, Takaaki Mano, T.~Ochiai, Stefano Sanguinetti, K.~Sakoda,
  G.~Kido, and Nobuyuki Koguchi.
\newblock {Optical transitions in quantum ring complexes}.
\newblock {\em Physical Review B}, 72(20):8, nov 2005.

\bibitem{PAVGUZ94}
L.~Pavesi and M.~Guzzi.
\newblock {Photoluminescence of
  AI{\$}{\_}{\{}x{\}}{\$}Ga{\$}{\_}{\{}1-x{\}}{\$}As alloys}.
\newblock {\em Journal of Applied Physics}, 75(10):4779, 1994.

\end{thebibliography}

\end{document}